# Room-Temperature Structures of Solid Hydrogen at High Pressures

Hanyu Liu, Li Zhu, Wenwen Cui and Yanming Ma*

*State Key Lab of Superhard Materials, Jilin University, Changchun 130012, China*

By employing first-principles metadynamics simulations, we explore the 300 K structures of solid hydrogen over the pressure range 150–300 GPa. At 200 GPa, we find the ambient-pressure disordered hexagonal close-packed (*hcp*) phase transited into an insulating *partially ordered hcp* phase (*po-hcp*), a mixture of ordered graphene-like $H_2$ layers and the other layers of weakly coupled, disordered $H_2$ molecules. Within this phase, hydrogen remains in paired states with creation of shorter intra-molecular bonds, which are responsible for the very high experimental Raman peak above 4000 cm$^{-1}$. At 275 GPa, our simulations predicted a transformation from *po-hcp* into the ordered molecular metallic *Cmca* phase (4 molecules/cell) that was previously proposed to be stable only above 400 GPa. Gibbs free energy calculations at 300 K confirmed the energetic stabilities of the *po-hcp* and metallic *Cmca* phases over all known structures at 220-242 GPa and >242 GPa, respectively. Our simulations highlighted the major role played by temperature in tuning the phase stabilities and provided theoretical support for claimed metallization of solid hydrogen below 300 GPa at 300 K.

**I. Introduction**

Hydrogen has attracted much attention because it is the simplest element, consisting of one electron and one proton, and is the most abundant element in the universe. In 1935, Wigner and Huntington firstly proposed that solid hydrogen might dissociate into an atomic metal around 25 Gpa.[1] Later, metallic hydrogen was predicted to be a good candidate for high-temperature superconductors.[2] Extensive high-pressure experimental and theoretical investigations of solid hydrogen have since been conducted.[3-12] The recently proposed high-pressure concept of a metallic superfluid[13] or a quantum liquid[14] for solid hydrogen has generated excitement in the field.

High-pressure structures of solid hydrogen are central to an understanding of the related physical properties. Unfortunately, the extremely weak X-ray scattering of hydrogen has hindered experimental studies of the structures of low-temperature and high-pressures but the low-pressure disordered *hcp* structure (phase I).[15] Other low-temperature and high-pressure phases II and III, discovered by vibrational spectroscopic experiments, have remained unsolved.[3,16-18] A variety of theoretical techniques have been employed to explore the structures of phases II and III (see, e.g., Refs. [6,10,12,19-21]); however, the interpretation of these studies has been extensively debated. The experimental findings[22] of the incommensurate nature of phase II of solid deuterium has introduced additional difficulties into the structural solution of phase II of solid hydrogen. The use of *ab initio* random structural searches led to the proposal of an energetically favorable monoclinic *C*2/*c* structure for phase III at zero temperature.[10] Encouragingly, the vibrational properties of the *C*2/*c* structure agreed somewhat with the experimental Raman data collected from phase III.[10] Theoretical predictions[10,23,24] of the structures up to terapascal pressures proposed structural models of metallic hydrogen in molecular or atomic forms. The theoretical studies reached an apparent consensus that the metallization of solid hydrogen should occur above about 400 GPa.[21,25] High-pressure spectroscopic studies in the search for metallic state of solid hydrogen have been extensively conducted (see, e.g., Refs. [3,5,8,9,11,16-18,26-28]). Experiments up to a highest pressure of 300 GPa at low temperatures (100 K) did not yield evidence for metallization.[9,28]

Recently, Eremets and Troyan[29] reported the room temperature experimental observation of metallic hydrogen at surprisingly low pressures of 260–270 GPa. Before entering the metallic phase, they also observed the stabilization of a new phase (phase IV) whose Raman spectrum was hardly interpreted by available structural models. This newly observed phase IV was also identified in another experiment[30] and interpreted as a mixed phase that consists of atomic hydrogen and unbound $H_2$ molecules which was earlier theoretically predicted as *Pbcn* structure.[10] Recent synchrotron infrared and optical absorption measurements suggested that the insulating broken-symmetry phase III with paired hydrogen states remained stable at least up to 360 GPa over a broad temperature range.[31] Soon after the experimental suggestion of the mixed *Pbcn* structure as a candidate for phase IV, an slightly improved *Pc* structure with more distorted graphene-like sheets was proposed by Pickard *et al.*[32] to eliminate most of the severe imaginary phonons of *Pbcn* structure.

These recent researches significantly advanced the field, but clearly raised two major questions: (i) what might be the true structure of phase IV, and (ii) whether or not metallic hydrogen could be made at such low pressures (<300 GPa)[33] and what is the metallic structure? An understanding of these important questions would require a thorough investigation of the high-pressure structures of solid hydrogen at *300 K*, at which temperature the experiments were performed. Here, we conducted extensive first-principles metadynamics simulations of solid hydrogen at 300 K and pressures of 150–300 GPa, in an effort to answer the above questions.

## II. THEORETICAL DETAILS

**Metadynamics simulations:**

The metadynamics algorithm[34,35] for the study of structural phase transitions was employed. This technique searches for low energy pathways from the initial energy well to neighboring minima and provides a means to simulate structural transformations in crystals. The metadynamics method is able to overcome barriers and therefore can explore a much broad range of candidate structures at finite temperatures. Successful applications of the method include several examples of reconstructive structural transitions (see, e.g., Refs [36-40]). In the present study, the metadynamics method was applied by a combination with the projector augmented plane-wave (PAW) method[41] as implemented in the Vienna *ab initio* Simulation

Package (VASP) code.[42] A PAW potential with a Perdew-Burke-Ernzerhof[43] exchange-correlation functional was adopted. The simulation cells were constructed by using 64 hydrogen molecules and Brillouin zone was sampled with the 4×4×4 k-points. The canonical *NVT* (*N*-number of particles, *V*-volume, *T*-temperature) ensemble was used for molecular dynamics runs. Each metastep of metadynamics simulations includes 600 time steps with each time step of 0.5 fs. Extensive metadynamics simulations with typical 100 metasteps for each simulation at 300 K and pressures of 150–300 GPa were conducted using Gaussian width $\delta = 50$ (kbar Å$^3$)$^{1/2}$ and Gaussian height $W = 2500$ kbar Å$^3$, which are typically good parameters for investigation of phase transitions.[35,38]

**Gibbs free energy calculations:**

To determine the thermodynamic stability of various phases at finite temperature, we have then carried out phonon calculations to obtain Gibbs free energies (*G*) for competitive structures using the harmonic approximation:

$$G = U + PV + \frac{1}{2}\sum_{q,j}\hbar\omega_{q,j} + K_B T \sum_{q,j} \ln\left\{1 - \exp\left[-\hbar\omega_{q,j}/K_B T\right]\right\} \quad (1)$$

Here $U$, $P$ and $V$ are the DFT total energy, volume and external pressure of the static structures, respectively. The third and fourth terms are zero-point motion and vibrational contributions to the free energy, respectively, and $\omega(q,j)$ is the phonon frequency of *j*th mode at wave vector *q* in the Brillouin zone. The phonon frequencies were calculated in supercells containing 48 molecules for *C2/c*, *Cmca*-4, *Cmca*-12, *Pc*, and *Cc* structures using FROPHO program.[44] A Brillouin zone sampling grid with 6×6×6 k-mesh, with a plane wave basis set cutoff of 500 eV. Note that in the Gibbs free energy (or enthalpy) vs. pressure calculations; the pressure is limited to the static DFT pressure. The estimated vibrational pressure is about 10 GPa at a static pressure of 250 GPa.

**Two phase simulations of melting temperature:**

We determine the melting temperature from the *ab initio* simulation with the *NPT* (*N*-number of particles, *P*-pressure, *T*-temperature) ensemble, as recently implemented in VASP code.[45] The simulation box was constrained to remain in a tetragonal shape, with the two short sides (parallel to the plane of the interface) being of equal length. 960 hydrogen molecules were used in the (6×4×10) simulation cell of the *Cmca*-4 phase. Initial configurations containing approximately equal amounts of solid and liquid plus an interface between them were generated. Our two phase simulations result in a single phase: above melting temperature $T_m$, the system will become liquid, whereas below $T_m$, the liquid part will solidify. The Brillouin zone was sampled with the Γ point. The time step in the molecular-dynamics simulations was 0.5 fs, and the self-consistency on the total energy was $2\times10^{-5}$ eV.

## III. Results and Discussion

At 150 GPa and 300 K, metadynamics simulations were performed with an initial structure of the ambient-pressure disordered phase I. No obvious structural changes were observed after 100 metastep runs, a typical long run to model the phase transitions as learnt from previous simulations.[37,38,46] This indicated that the initial structure of phase I remained, in good agreement with the experimental results.[5,26,29,30] Upon compression to 200 GPa, after 44 metastep runs, phase I was no longer persistent, and an *partially ordered* new structure was stabilized (inset of Fig. 1a). This structure at 300 K was consistent with an *hcp* lattice in which the centers of $H_2$ molecules deviated slightly from the ideal *hcp* lattice. The simulated structure is thus seen as a *partially ordered hcp* (*po-hcp*) structure, which might be related to either experimentally observed phase I'[17] or IV[29,30]. Careful inspection of *po-hcp* structure yielded a mixed phase which includes two different molecular layers: (i) layers of weakly coupled $H_2$ molecules (layer I) with shorter intra-molecular bonds ~0.73 Å, and (ii) strongly coupled $H_2$ molecules forming graphene-like planar sheets (layer II) with longer intra-molecular bondlengths ~0.78 Å. Our structure shows explicitly similar to the ordered *Pbcn* or *Pc* structures proposed by Pickard *et al.*[10,32] However, our simulations were able to identify that $H_2$ molecules in layer I are orientationally *disordered* (Fig. 2 *a* and *b*), where the distribution of protons is spherical without showing any orientational order. In layer II, in contrast, the distribution of protons indicates the formation of *ordered* molecular graphene-like sheets (Fig. 2c). It is noteworthy that existence of two types of longer and shorter intra-molecular bonds (Fig. 3) in this *po-hcp* structure was critical for interpreting the Raman spectra[29,30] of the newly observed phase IV at 300 K. Creation of shorter bonds explained well the experimental observation[29,30] of a very high-frequency Raman peak above 4000 cm$^{-1}$. Larger increases of longer bonds upon compression (Fig. 3) are in good accordance with the experimental observations[29,30] of a marked decrease in the vibronic frequencies at 300 K.

The disordered nature of $H_2$ molecules in layer I is not unreasonable since the inter-molecular distance is ~1.31 Å at 200 GPa, evidently larger than that (1.23 Å) in completely disordered phase I at 150 GPa. This allows relatively looser $H_2$ environment, making the $H_2$ disorder possible. Subsequently, a particularly designed total-energy calculation was performed through frozen orientation technique, i.e., gradually rotating $H_2$ molecules of layer I in (100) and ($1\bar{1}0$) planes while remaining the centers of $H_2$ unchanged. It was found that the maximal energy barriers of $H_2$ rotations are ~32 meV per $H_2$ at 200 GPa (Fig. 4). Such a low-energy barrier could naturally lead to a free rotator of $H_2$, resulting in an orientational disorder. For a direct comparison, we have conducted the similar rotational energy barrier calculations on phase I at 150 GPa (Fig. 4). The resulting maximal energy barrier is ~28 meV per $H_2$ at 150 GPa, very close to that calculated for our *po-hcp* structure.

Because the *po-hcp* structure at 200 GPa is partially disordered, it is difficult to calculate the vibrational spectrum, and thus the Gibbs free energy. It is therefore necessary to extract an approximately ordered structure for the subsequent calculations of physical properties. We thus performed additional room-temperature *NPT*-MD (*N*-number of particles, *P*-pressure, *T*-temperature, MD-molecular dynamics)[45,47] simulations by adopting a much larger simulation cell of 288 molecules. Brillouin zone was sampled by 2×2×2 k-points. The fully optimized structures at 300 K were then relaxed at 0 K. The resultant ordered approximate structure has *Cc* symmetry with 96 molecules per unit cell (Table I) and was used for the subsequent discussions. We note that this *Cc* structure does not contain any imaginary phonons as confirmed by our calculations of phonon dispersion curves (Fig. 5).

At higher pressures of 275 GPa and 300 K, metadynamics simulations were performed using the initial simulation cell constructed from the *po-hcp* phase. After 26 metastep runs, the structure underwent a remarkable reconstruction with stabilization of a molecular orthorhombic structure (inset of Fig. 1*b*), which could no longer be described as having hexagonal symmetry. The resultant structure was consistent with the *Cmca*-4 structure (4 molecules per unit cell), which has been proposed in other low-temperature calculations.[25,48-50] Notably, Kohn–Sham density functional theory (DFT) previously predicted that this structure was stable merely above 400 GPa,[10,21] Here, our metadynamics simulations at 300 K lowered the pressure range associated with a stable *Cmca*-4 structure to below 300 GPa. This illustrated the important role of temperature played in tuning structural stability and indicated that lattice energy must be considered in the understanding of phase stability of solid hydrogen.

At 0 K, Gibbs free energy (G) reduces to enthalpy + zero-point energy (ZPE). Hydrogen has the lightest atomic mass, and therefore its ZPE can already be large enough to affect the structural stabilities. Indeed, ZPEs of *C*2/*c*, *Cmca*-4, *Cmca*-12 (12 molecules/cell, predicted in Ref. [10]), *Cc* and *Pc* structures at 250 GPa are calculated to having very large values of 605, 585, 604, 600, 599 meV/molecule, respectively. The computed enthalpies of various structures in the pressure range 200–300 GPa with and without the inclusion of ZPE at 0 K are shown in Fig. 6 a and b, respectively. It is found that the *Cc* and *Pc* structures are energetically degenerate in a large pressure range as expected. Within the static-lattice approximation, it is clearly seen that the *C2/c* structure transformed into *Cmca*-12 phase at 295 GPa (Fig. 6*a*), in good agreement with previous calculations.[10] However, once ZPE was considered, the *Cmca*-12 structure is never stable in all pressures studied; instead *Cmca*-4 structure becomes energetically more favorable than *C2/c* structure at a surprisingly low pressure of 237 GPa (Fig. 6*b*), though at such a low pressure, experiments do not observe any obvious phase transition. At an elevated temperature of 300 K, both *Cc* and *Pc* structures entered the stable pressure range at 220-242 GPa (Fig. 6*c*), illustrating the importance of temperature effects. This also demonstrates a fundamental feature that po-hcp phase is a thermally driven partially disordered phase. Our Gibbs free energy calculations revised significantly the earlier phase stability picture of solid hydrogen established at zero temperature, and confirmed our metadynamic simulations on the 300 K stabilization of *Cmca*-4 structure.

We examined the electronic properties of the *C*c and *Cmca*-4 structures at 250 GPa by calculating their electronic band structures and DOS (Fig. 7). The results suggested that *Cc* structure is a semiconductor with a band gap of 1.2 eV. This value was expected to be larger since exact-exchange DFT calculations demonstrated that the true bandgaps in hydrogen solids are roughly 1–2 eV higher than the standard DFT values.[21] The *Cmca*-4 structure is a weak metal with a low DOS at the Fermi level (Fig. 7b). It is known that the screened hybrid functional of Heyd, Scuseria, and Ernzerhof (HSE)[51] has the ability to accurately predict electronic band gaps. We thus use HSE functional to correct the possible band gap errors derived from the standard DFT calculations, and the results confirmed that *Cmca*-4 structure is a true metal (Fig. 7c). This electronic feature clearly agreed with the experimental observation of the metallization of solid hydrogen at 260–270 GPa,[29] but in contrast to other experiments.[30,31]

The harmonic approach was further used to construct the phase boundary of hydrogen (Fig. 8). Remarkably, the two experimentally observed 300 K phases[29] (solid circle for the insulating phase and solid square for the metallic phase in Fig. 4) agreed well with the simulated phase regimes of the *Cc* and *Cmca*-4 structures (Fig. 8). Notably, our phase diagram predicted that metallization should also occur below 300 GPa at low temperatures. This prediction appeared to contradict earlier experimental observations of a non-metallized phase at 300 GPa and 100 K.[9,28] The discrepancy may stem from the existence of a large kinetic barrier at the formation of metallic phase. Direct calculation of the enthalpy as a function of the reaction coordinate shows that the kinetic barrier is very large (0.246 eV/molecule) (Fig. 9). For comparison, the large kinetic barrier for the graphite→diamond transition is 0.23 eV/atom at 15 GPa.[52] where a much higher temperature (~1300 K) is needed to initiate the transition. Experimentally, once the metallic hydrogen had formed at 300 K, it remained stable upon cooling to at least 30 K.[29]

Previous calculations provided evidence that solid hydrogen melts into a metallic liquid above 300 GPa and 400 K.[53] Thus, it is unclear whether the observed conductive phase is a metallic liquid. We ruled out this possibility by performing additional simulations. A reliable two-phase (liquid and solid) approach[45,54] was adopted to determine the melting temperature of the *Cmca*-4 structure at 300 GPa by using a simulation cell of 960 hydrogen molecules. The melting temperature was predicted to be 580 ± 50 K, well above 300 K. As another support on the validity of *Cmca*-4 structure, at 300 GPa we heated solid hydrogen up to 1000 K and as expected solid hydrogen completely melts. The resultant liquid hydrogen was then quenched to 300 K and after 2 ps simulations of molecular dynamics, the liquid product was remarkably re-crystallized in the *Cmca*-4 structure, which is in excellent coincidence with our above metadynamic simulations of a stable 300 K *Cmca*-4 structure.

## IV. CONCLUSIONS

In summary, we explored the 300 K crystal structures of solid hydrogen over the pressure range 150–300 GPa using first-principles metadynamics methods. Our simulations modeled the transformation of phase I in turn into insulating *po-hcp* and metallic *Cmca*-4 phases at 300 K. Our simulations of phase IV do not contradict earlier structural models, but reveal a surprisingly *partially ordered* structural scenario: $H_2$ molecules are ordering in one layer, but disordering in alternate layer. This is the first example established thus far for a *partially ordered* structural feature among molecular solids. We demonstrated that temperature effects are critically important to an understanding of the high-pressure behavior of solid hydrogen. Our simulations provide theoretical evidence on the feasible metallization of solid hydrogen at 300 K below 300 GPa, in agreement with a recent 300 K experiment[29], but having discrepancy with other experiments.[30,31] Further experimental and theoretical studies on solid hydrogen at 300 K are greatly stimulated to understand this discrepancy.

**Acknowledgments:**

The authors acknowledge the High Performance Computing Center of Jilin University for supercomputer time, Changjiang Scholar and Innovative Research Team in University (No. IRT1132), funding from the China 973 Program under Grant No. 2011CB808200, and National Natural Science Foundation of China under grant Nos. 11025418 and 91022029.

*Author to whom correspondence should be addressed: mym@jlu.edu.cn

**Table I.** Structural parameters for the *Cc* structure at 250 GPa.

| | Lattice Parameters (Å, deg) | | Atomic Coordinates (Fractional) | | |
|---|---|---|---|---|---|
| *Cc* | a=b=5.8089 | H1 | 0.58415 | 0.25245 | 0.82235 |
| | c=5.1671 | H2 | 0.95937 | 0.62051 | 0.5683 |
| | α = β =89.99 | H3 | 0.73174 | 0.39919 | 0.31973 |
| | γ =119.92 | H4 | 0.65738 | 0.53365 | 0.57421 |
| | | H5 | 0.90231 | 0.72976 | 0.8219 |
| | | H6 | 0.77621 | 0.61455 | 0.85439 |
| | | H7 | 0.86987 | 0.83515 | 0.56996 |
| | | H8 | 0.61168 | 0.27901 | 0.35714 |
| | | H9 | 0.52538 | 0.45482 | 0.32566 |
| | | H10 | 0.6551 | 0.0344 | 0.57051 |
| | | H11 | 0.311 | 0.99315 | 0.07121 |
| | | H12 | 0.53344 | 0.15675 | 0.0771 |
| | | H13 | 0.68016 | 0.17982 | 0.07224 |
| | | H14 | 0.68238 | 0.68203 | 0.57709 |
| | | H15 | 0.83129 | 0.37044 | 0.06817 |
| | | H16 | 0.81112 | 0.49738 | 0.56912 |
| | | H17 | 0.6159 | 0.45717 | 0.07253 |
| | | H18 | 0.49275 | 0.31020 | 0.07199 |
| | | H19 | 0.99372 | 0.80986 | 0.56745 |
| | | H20 | 0.80797 | 0.49431 | 0.0656 |
| | | H21 | 0.8302 | 0.11587 | 0.84879 |
| | | H22 | 0.17565 | 0.18034 | 0.57112 |
| | | H23 | 0.53561 | 0.46294 | 0.81791 |
| | | H24 | 0.37031 | 0.33004 | 0.5663 |
| | | H25 | 0.86897 | 0.3338 | 0.56829 |
| | | H26 | 0.82665 | 0.23763 | 0.82627 |
| | | H27 | 0.75447 | 0.15891 | 0.31522 |
| | | H28 | 0.87845 | 0.16005 | 0.29422 |
| | | H29 | 0.3723 | 0.20651 | 0.28963 |
| | | H30 | 0.955 | 0.11912 | 0.57147 |
| | | H31 | 0.45195 | 0.11433 | 0.57505 |
| | | H32 | 0.49462 | 0.30902 | 0.57049 |
| | | H33 | 0.52279 | 0.95237 | 0.32244 |
| | | H34 | 0.25256 | 0.08195 | 0.31988 |
| | | H35 | 0.46508 | 0.03752 | 0.32031 |
| | | H36 | 0.56678 | 0.02736 | 0.82739 |

| | | | |
|---|---|---|---|
| H37 | 0.15511 | 0.03513 | 0.57379 |
| H38 | 0.42324 | 0.95886 | 0.81836 |
| H39 | 0.18103 | 0.67776 | 0.07424 |
| H40 | 0.99553 | 0.80868 | 0.07124 |
| H41 | 0.32642 | 0.23601 | 0.82258 |
| H42 | 0.32969 | 0.11541 | 0.85346 |
| H43 | 0.92223 | 0.9564 | 0.81837 |
| H44 | 0.37417 | 0.70461 | 0.28667 |
| H45 | 0.02839 | 0.06576 | 0.32377 |
| H46 | 0.30685 | 0.99043 | 0.57427 |
| H47 | 0.65689 | 0.75583 | 0.82207 |
| H48 | 0.66163 | 0.87892 | 0.78687 |

**Figure captions**

**Fig. 1**: (**color online**) Enthalpy evolution during the 300 K metadynamics simulations on transitions from the disordered *hcp* phase I to the *po-hcp* mixed phase at 200 GPa (a) and from *po-hcp* phase to the *Cmca*-4 phase at 275 GPa (b). The insets of (a) and (b) show the atomic configurations at lowest enthalpies at 200 GPa and 275 GPa, respectively.

**Fig. 2:** (**color online**) (a) Distributions of protons are extracted from the trajectory data of 3 ps molecular dynamics at 200 GPa and 300 K. Top views of (b) disordered and (c) ordered $H_2$ layers, respectively.

**Fig. 3:** (**color online**) The intra-molecular bond-lengths in various phases (*Cc*, *Pc*, *C2/c* and *hcp* structures) are plotted as a function of pressure.

**Fig. 4:** (**color online**) The illustration of $H_2$ rotations in the (a) (100) and (b) ($1\bar{1}0$) planes of *Cc* structure. The molecules with black spheres are counter-clockwisely rotated while the molecules with white spheres are fixed. (c) and (d) are the total-energies of *Cc* structure at 200 GPa and *hcp* structure at 150 GPa as a function of rotational angle, respectively. Inset of (d) represents $H_2$ rotations for *hcp* phase. The lowest energies are scaled as the zero energy.

**Fig. 5:** (**color online**) Phonon dispersions of the *Cc* structure at 250 GPa.

**Fig. 6:** (**color online**) Enthalpies/free energies of the *Cc*, *Pc*, *Cmca*-4 and *Cmca*-12 structures relative to *C*2/*c* structure (a) for the static lattices, (b) with ZPE, and (c) with full vibrational motion at 300 K.

**Fig. 7:** The calculated band structures and DOSs (in units of eV$^{-1}$ per molecule) for (a) *Cc* and (b) *Cmca*-4 structures with Perdew-Burke-Ernzerhof (PBE) functional at 250 GPa, respectively. In order to accurately predict electronic band gaps of *Cmca*-4 structure, we have calculated band structures and DOS (in units of eV$^{-1}$ per molecule) for the *Cmca*-4 structure with Heyd-Scuseria-Erhzerhof (HSE) functional at same pressure. It is clearly seen that the *Cmca*-4 structure is a weak metal with a low DOS at the Fermi level.

**Fig. 8:** (**color online**) Phase diagram of hydrogen. The red dashed lines between III, IV and V were calculated using a harmonic approach by assuming *C*2/c, *Cc* and *Cmca*-4 structural models, respectively. The open diamond represents melting temperature of *Cmca*-4 structure at 300 GPa calculated by using first-principles two-phase approach. The solid circle and square denoted the experimental pressures at which the sample transited to the newly observed insulating and metallic phases of solid hydrogen, respectively.[29] The boundaries between phases I–II–III at low temperatures were taken from Refs [16,30]. At higher temperatures, the melting curve was plotted according to the calculations in Ref [55]. Another theoretical melting data in Ref [14] are represented by open triangle. The experimental melting data in Ref [56] are indicated by solid vertical bars. The calculated phase boundary between the molecular and monatomic metallic liquids is shown above the melting line.[55] The open black star indicates liquid hydrogen, as predicted by quantum Monte Carlo simulations.[53]

**Fig. 9:** Activation barrier for the $C2/c \rightarrow Cmca$-4 transition at 250 GPa. Along the reaction coordinate, the hessian-matrix is derived by additional metadynamics simulations and changes from the values characteristic of $C2/c$ to those of $Cmca$-4. At each point all atomic positions were fully optimized under constraint of fixed hessian-matrix.

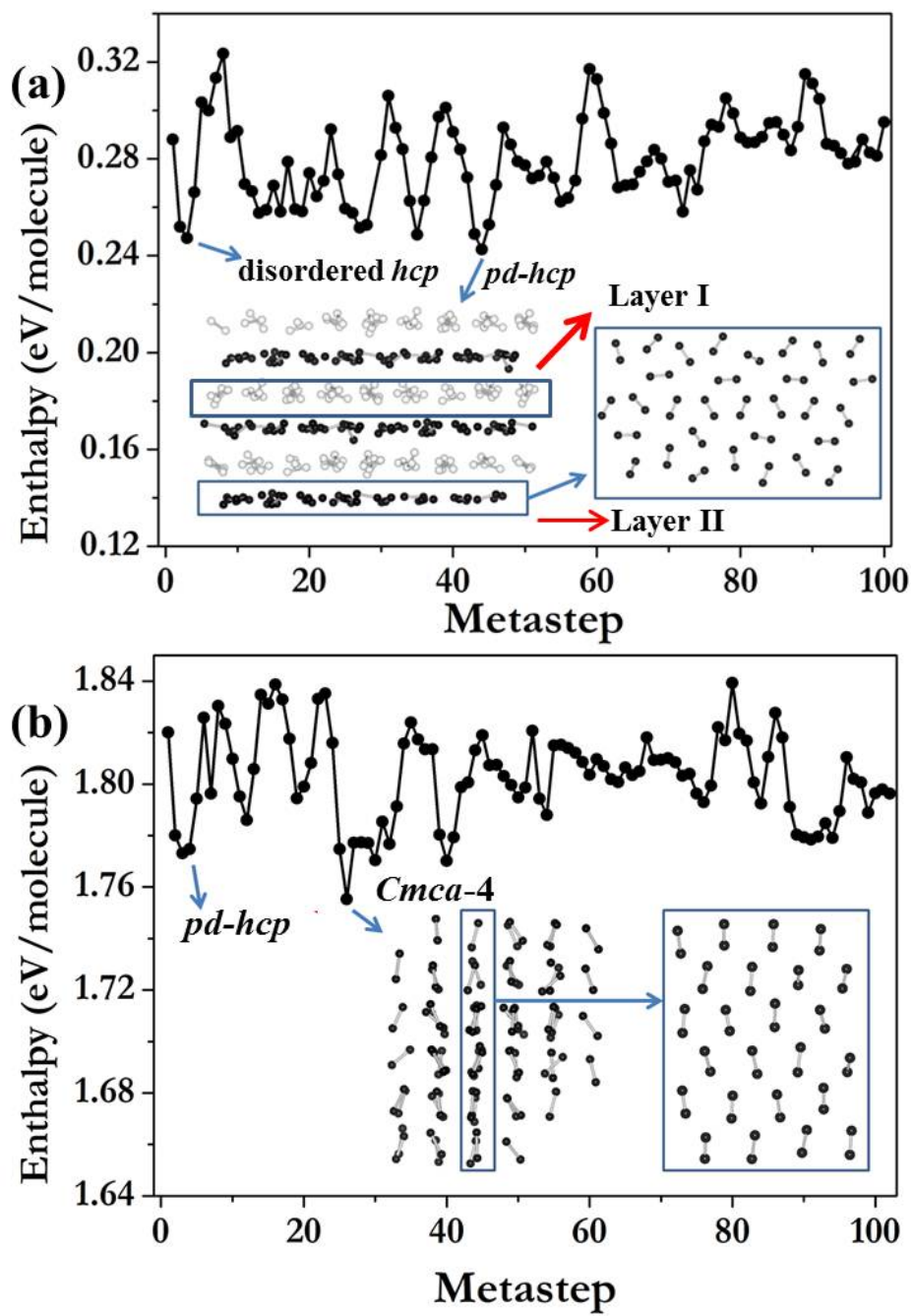

**Fig. 1**

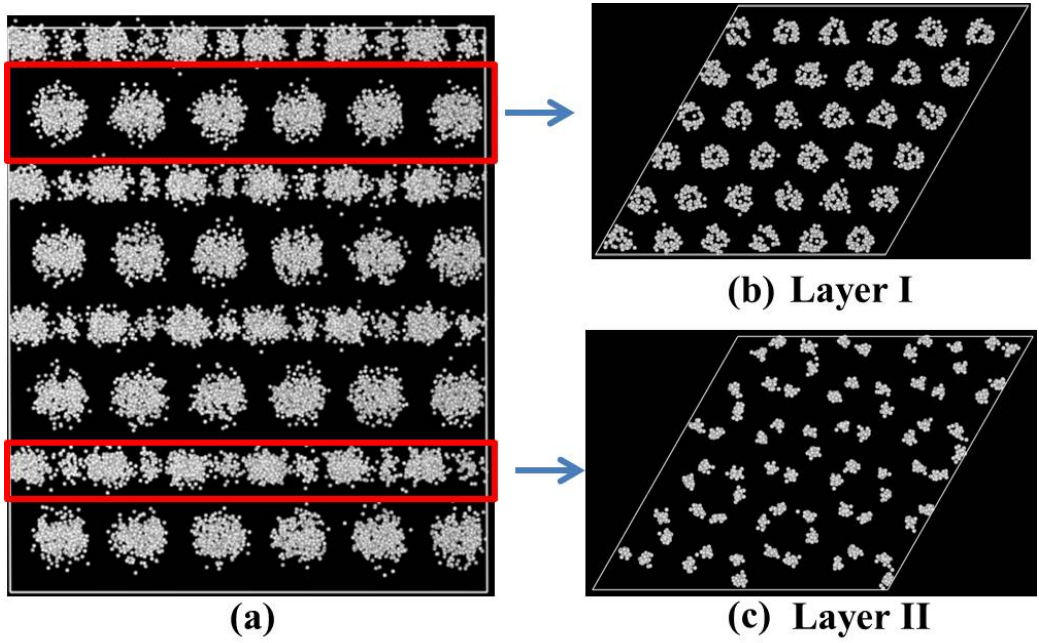

**Fig. 2**

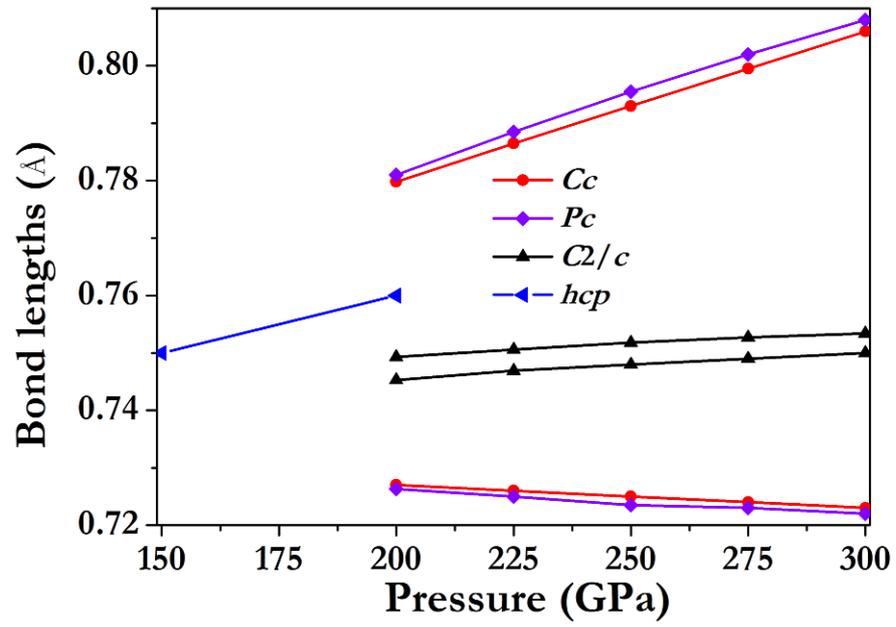

**Fig. 3**

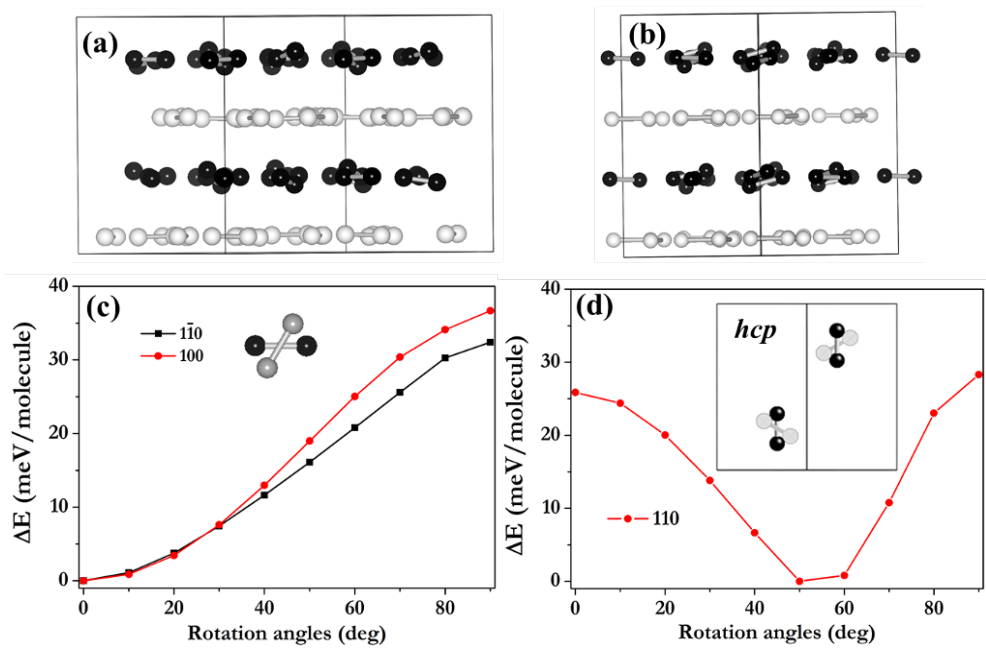

**Fig. 4**

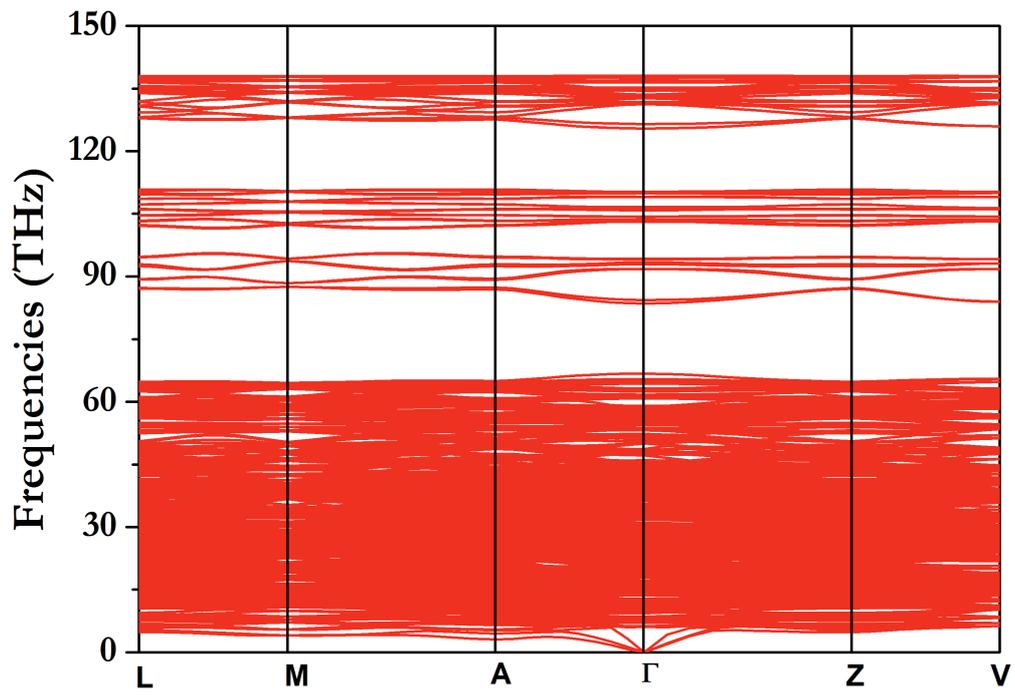

**Fig. 5**

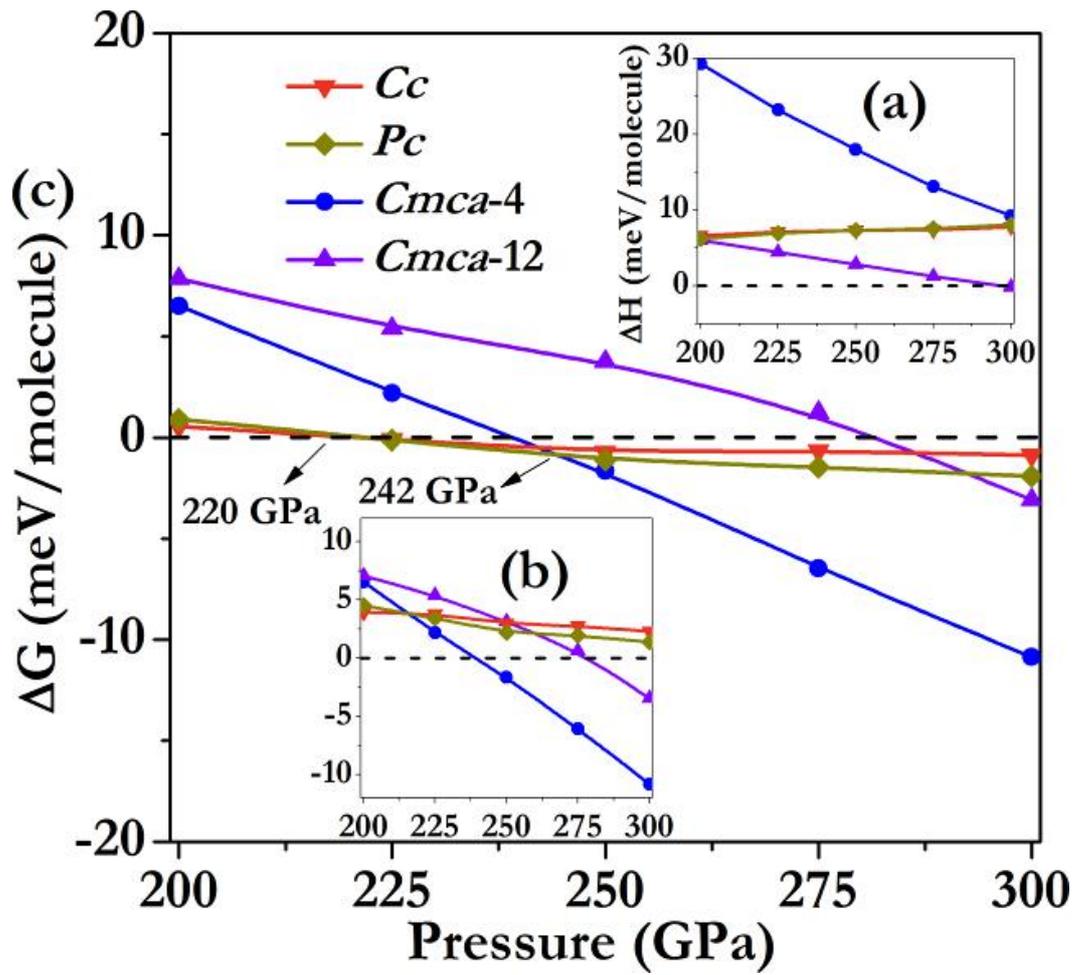

**Fig. 6**

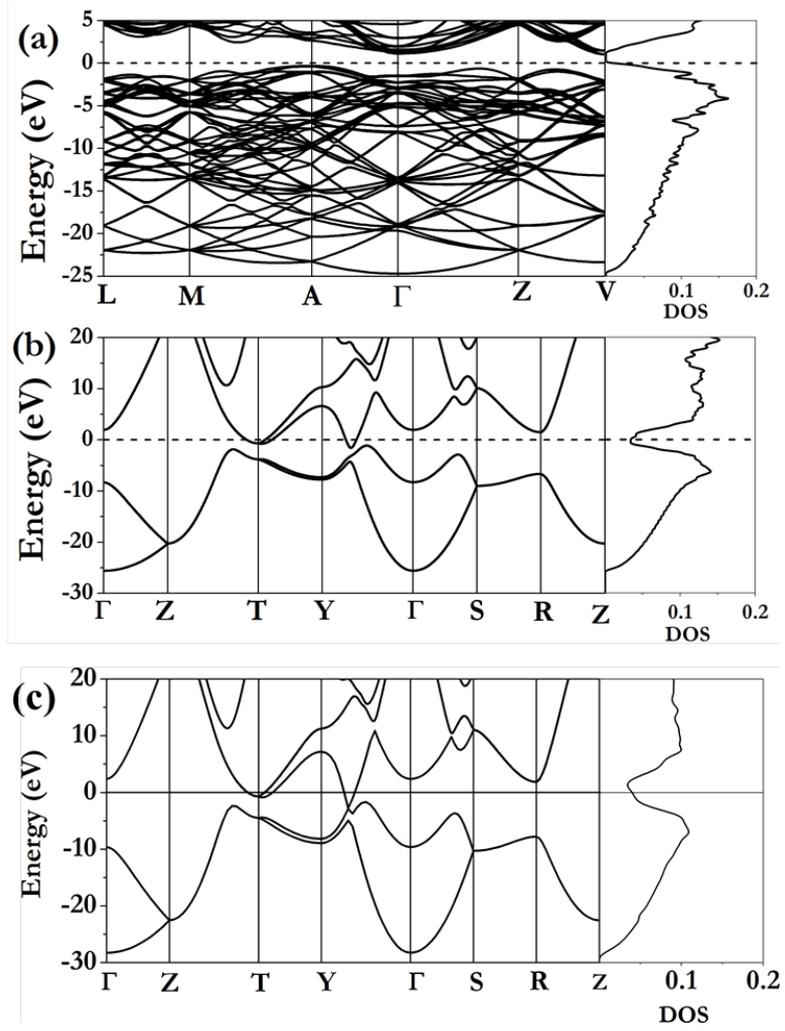

Fig. 7

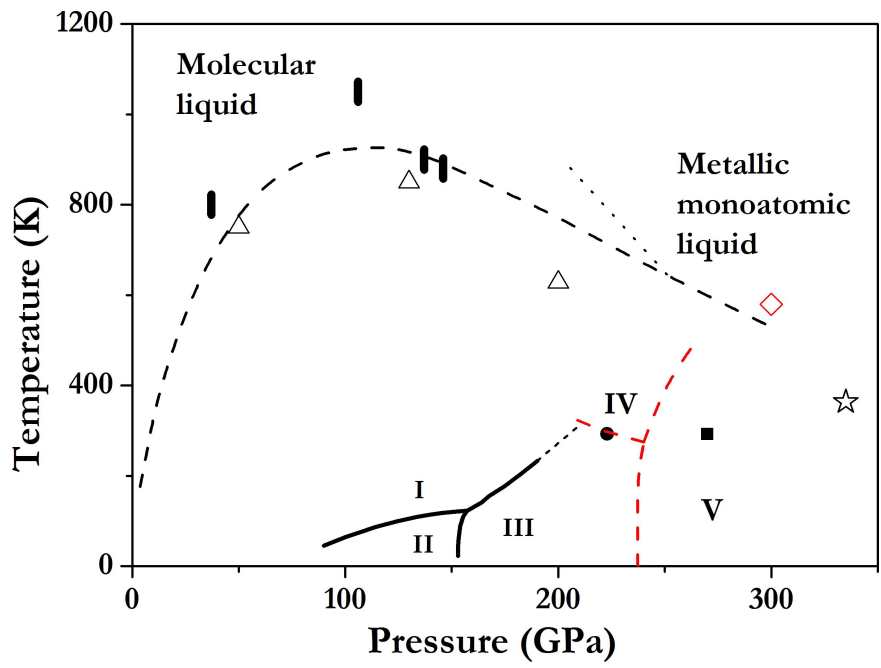

**Fig. 8**

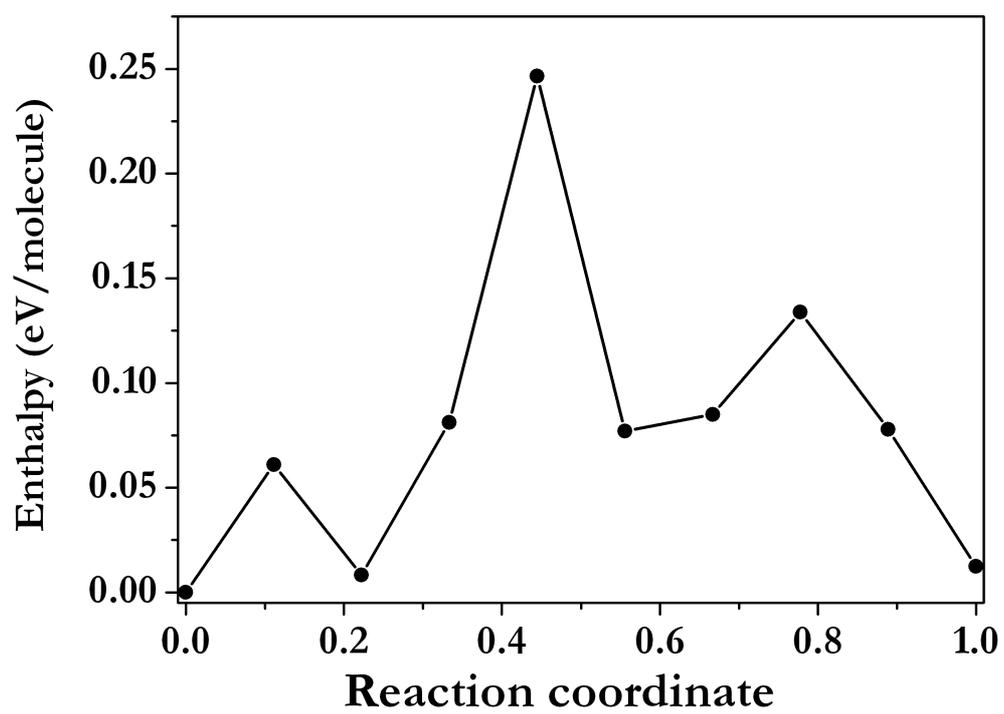

Fig. 9